\documentclass[12pt]{article}
\begin{document}
\begin{center}
{\large \bf ON CP-ODD EFFECTS IN $ K_L \to 2\pi$ AND  $K^{\pm} \to
\pi^{\pm} \pi^{\pm} \pi^{\mp} $ DECAYS GENERATED BY
DIRECT CP VIOLATION}\\ \vspace{5mm}  E.P. Shabalin \footnote{E-mail:
shabalin@heron.itep.ru}\\ Institute for Theoretical and Experimental
Physics, Moscow, Russia \end{center} \vspace{3cm} \begin{abstract} The
amplitudes of the $K^{\pm} \to 3\pi$ and $K \to 2\pi$ decays are expressed
in terms of different combinations of one and the same set  of
CP-conserving and CP-odd parameters. Extracting the magnitudes of these
parameters from the data on $K \to 2\pi$ decays, we estimate an expected
CP-odd difference between the values of the slope parameters $g^+$ and
$g^-$ of the energy distributions of "odd" pions in $K^+ \to \pi^+\pi^+
\pi^-$  and $K^- \to \pi^-\pi^- \pi^+$ decays.  \end{abstract} \newpage
\section*{1. Introduction}

The observation of CP effects in $K^{\pm} \to 3\pi$
decays would allow to understand better how the mechanisms of CP violation
work.

Now the Collaboration NA48/2  began a search for such effect
with accuracy $\delta(\frac{g^+ -g^-}{g^+ +g^-})\le 2\times 10^{-4}$.

Contrary to the case of $K_L \to 2\pi$ decay, where CP violates both in
$\Delta S =2$ and $\Delta S=1$ transitions, in the $K^{\pm} \to 3\pi$
decays, only the last (the so-called "direct" ) CP violation takes place.
Experimentally, an existence of the direct CP violation in $ K_L \to 2\pi$
decays, predicted by Standard Model (SM) and characterized by the
parameter $\varepsilon'$, is establised: $\varepsilon'/\varepsilon =(1.66
\pm 0.16)\times 10^{-3}$.

What is expected for CP effects in $K^+ \to \pi^{\pm} \pi^{\pm} \pi^{\mp}$
decay? To give an answer, it is necessary to understand the role of the
electroweak penguin (EWP) operators in both decays and get rid of the
large uncertainties usual for the theoretical calculations. The real scale
of these uncertainties is characterized by the following predictions
obtained before the above experimental result:

$$ \frac{\varepsilon'}{\varepsilon}=(17^{+14}_{-10})\times 10^{-4} \quad
[1], \qquad  \frac{\varepsilon'}{\varepsilon}=(1.5\div 31.6)\times 10^{-4}
\quad [2] $$ .
 To avoid the uncertainties arising in the theoretical calculation of the
ingredients of the theory, we use the following procedure. We
express the amplitudes of $K \to 2\pi$ and $K^{\pm} \to 3\pi$ decays in
terms of one and the same set of parameters, and calculating $g^+ -g^-$, we
use the magnitudes of these parameters extracted from data on $K \to
2\pi$ decays.
\section*{2. The scheme of calculation}

A theory of $\Delta S=1$ non-leptonic decays is based on the effective
Lagrangian [3]
\begin{equation}
L(\Delta S=1) =\sqrt{2}G_{\rm F}\sin\theta_{\rm C} \cos \theta_{\rm C}
\sum_i c_iO_i\,, \label{1} \end{equation} where \begin{equation} O_1=\bar
s_L \gamma_{\mu}d_L \cdot \bar u_L \gamma_{\mu} u_L - \bar s_L
\gamma_{\mu} u_L \cdot \bar u_L \gamma_{\mu} d_L \qquad(\{8_f\}, \Delta I
= 1/2); \end{equation} \begin{eqnarray} O_2=\bar s_L \gamma_{\mu} d_L
\cdot \bar u_L \gamma_{\mu} u_L + \bar s_L \gamma_{\mu} u_L \cdot \bar u_L
\gamma_{\mu} d_L +2\bar s_L \gamma_{\mu} d_L \cdot \bar d_L \gamma_{\mu}
d_L \nonumber  \\ \qquad {}+2 \bar s_L \gamma_{\mu} d_L \cdot \bar s_L
\gamma_{\mu} s_L \qquad (\{8_d\}, \Delta I =1/2); \end{eqnarray}
\begin{eqnarray}
O_3=\bar s_L \gamma_{\mu} d_L \cdot \bar u \gamma_{\mu} u_L +\bar s
\gamma_{\mu} u_L \cdot \bar u \gamma_{\mu} d_L + 2 \bar s_L \gamma_{\mu}
d_L \cdot \bar d_L \gamma_{\mu} d_L \nonumber \\
\qquad {}-3 \bar s_L \gamma_{\mu} d_L \cdot \bar s_L \gamma_{\mu} s_L
\qquad (\{27\}, \Delta I=1/2);
\end{eqnarray}
\begin{eqnarray}
O_4=\bar s_L \gamma_{\mu} d_L \cdot \bar u \gamma_{\mu} u_L + \bar s_L
\gamma_{\mu} u_L \cdot \bar u_L \gamma_{\mu} d_L - \nonumber \\
\qquad {}-\bar s_L \gamma_{\mu} d_L \cdot \bar d_L \gamma_{\mu} d_L
\qquad (\{27\}, \Delta I=3/2);
\end{eqnarray}
\begin{equation}
O_5= \bar s_L \gamma_{\mu} \lambda^a d_L(\sum_{q=u,d,s} \bar q_R
\gamma_{\mu} \lambda^a q_R) \qquad (\{8\},  \Delta I=1/2);
\end{equation}
\begin{equation}
O_6= \bar s_L \gamma_{\mu} d_L (\sum_{q=u,d,s} \bar q_R \gamma_{\mu} q_R)
\qquad (\{8 \}, \Delta I = 1/2).
\end{equation}
This set is sufficient for calculation of the CP-even parts of the
amplitudes under consideration. To calculate the CP-odd parts , it is
necessary to add the so-called electroweak contributions originated by the
operators $O_7, O_8$:
\begin{equation}
O_7=\frac{3}{2} \bar s\gamma_{\mu}(1+\gamma_5)d (\sum_{q=u,d,s}e_q
\bar q \gamma_{\mu}(1-\gamma_5)q) \qquad  (\Delta I=1/2, 3/2);
\end{equation} \begin{equation} O_8=-12\sum_{q=u,d,s} e_q (\bar s_L
q_R)(\bar q_R d_L),\quad e_q=(\frac{2}{3}, -\frac{1}{3}, -\frac{1}{3})
\quad (\Delta I=1/2, 3/2). \end{equation} The coefficients $ c_{5-8}$ have
the imaginary parts necessary for CP violation.

In the case of non-linear realization of chiral symmetry, the bosonization
of these operators can be done using the relations [4]
\begin{equation} \bar
q_j(1+\gamma_5)q_k=-\frac{1}{\sqrt{2}}F_{\pi}r\left(U-\frac{1}{\Lambda^2}
\partial^2 U \right)_{kj} \, ,
\end{equation}
\begin{equation}
\bar q_j \gamma_{\mu}(1+\gamma_5) q_k=i[\left(\partial_{\mu}U
\right)U^{\dag} - U \left( \partial_{\mu} U^{\dag}  \right)-
\frac{rF_{\pi}}{\sqrt{2}\Lambda^2} \left(m(\partial_{\mu} U^{\dag}
)-( \partial_{\mu} U)m \right)]_{kj}
\end{equation}
where $$F_{\pi}\approx 93\, \mbox{MeV}, \quad \Lambda \approx
1\,\mbox{GeV},\quad r=2m^2_{\pi}/(m_u+m_d),\quad
m=\mbox{diag}(m_u,m_d,m_s).$$ \begin{equation}
U=\frac{F_{\pi}}{\sqrt{2}}\left(1+\frac{i\sqrt{2} \hat
\pi}{F_{\pi}}-\frac{\hat \pi^2}{F_{\pi}^2} +a_3\left(\frac{i\hat
\pi}{\sqrt{2}F_{\pi}}\right)^3 +2(a_3-1)\left(\frac{i\hat
\pi}{\sqrt{2}F_{\pi}}\right)^4 +...\right),
\end{equation}
where $a_3$ is an arbitrary number and
\begin{equation}
\hat \pi=
\left(
\begin{array}{lll}
\frac{\pi_0}{\sqrt{3}}+\frac{\pi_8}{\sqrt{6}}+\frac{\pi_3}{\sqrt{2}} &
\quad \pi^+ & \quad K^+ \\
\pi^- & \quad
\frac{\pi_0}{\sqrt{3}}+\frac{\pi_8}{\sqrt{6}}-\frac{\pi_3}{\sqrt{2}} &
\quad K^0 \\ K^- & \quad \bar K^0 & \quad
\frac{\pi_0}{\sqrt{3}}-\frac{2\pi_8}{\sqrt{6}} \end{array} \right)
\end{equation}
The PCAC condition demands $a_3=0$  [5] and we adopt this condition,
bearing in mind that, on mass shell, the values of the mesonic amplitudes
are independent of $a_3$.

Using also the relations between matrices in the color space
$$
\begin{array}{lll}
\delta^{\alpha}_{\beta} \delta^{\gamma}_{\delta} =\frac{1}{3}
\delta^{\alpha}_{\delta}
\delta^{\gamma}_{\beta}+\frac{1}{2}{\bf \lambda}^{\alpha}_{\delta}
{\bf \lambda}^{\gamma}_{\beta} \\
{\bf \lambda}^{\alpha}_{\beta}
{\bf \lambda}^{\gamma}_{\delta}=\frac{16}{9}\delta^{\alpha}_{\delta}
\delta^{\gamma}_{\beta} -\frac{1}{3} {\bf \lambda}^{\alpha}_{\delta}
{\bf \lambda}^{\gamma}_{\beta}
\end{array}
$$
and the Fierz transformation relation
$$
\bar s \gamma_{\mu}(1+\gamma_5) d \cdot \bar q \gamma_{\mu}(1-\gamma_5)q=
-2\bar s(1-\gamma_5)q \cdot \bar q(1+\gamma_5)d
$$
and representing $M(K \to 2\pi)$ in the form
\begin{eqnarray}
M(K^0_1 \to \pi^+\pi^-)=A_0e^{i\delta_0}-A_2e^{i\delta_2}, \\
M(K^0_1 \to \pi^0\pi^0)=A_0e^{i\delta_0}+2A_2e^{i\delta_2}, \\
M(K^+ \to \pi^+\pi^0) =-\frac{3}{2} A_2 e^{i\delta_2},
\end{eqnarray}
where $\delta_0$ and $\delta_2$ are the $S$ -wave shifts of $\pi\pi$
scattering in isotopic spin $I=0,2$ channels, we obtain \begin{eqnarray}
A_0=G_{\rm F}F_{\pi}\sin \theta_{\rm C}
\cos\theta_{\rm C}\frac{m^2_K-m^2_{\pi}}{\sqrt{2}} [c_1-c_2-c_3
+\frac{32}{9}\beta(\mbox{Re} \tilde c_5 +i \mbox{Im} \tilde c_5)]; \\
A_2=G_{\rm F}F_{\pi}\sin \theta_{\rm C} \cos \theta_{\rm C}
\frac{m^2_K-m^2_{\pi}}{\sqrt{2}} \cdot [c_4 +i \frac{2}{3}\beta \Lambda^2
\mbox{Im}\tilde c_7(m^2_K-m^2_{\pi})^{-1}] \end{eqnarray} where $$ \tilde
c_5=c_5+\frac{3}{16} c_6,\quad \tilde c_7=c_7+3c_8.  $$

\begin{equation}
\beta=\frac{2m^4_{\pi}}{\Lambda^2 (m_u+m_d)^2}.
\end{equation}
The contributions from $\tilde c_7 O_7$ into $\mbox{Re}A_0$ and
$\mbox{Im}A_0$ are small because $\tilde c_7/\tilde c_5 $ is proportional
to the electromagnetic constant $ \alpha$ and we neglected
these corrections.  From data on widths of $K \to 2\pi$ decays we obtain
\begin{equation} c_4 =0.328;\quad c_1-c_2-c_3+\frac{32}{9} \beta \mbox{Re}
\tilde c_5 =-10.13.  \end{equation} At $c_1-c_2-c_3=-2.89$ \quad [3], [6]
we obtain \begin{equation} \frac{32}{9} \beta \mbox{Re} \tilde c_5 = -
7.24.  \end{equation} From the expression for $A_2$, it is seen that the
contribution of the operators $O_{7,8}$  is enlarged by the factor
$\Lambda^2/m^2_K$ in comparison with the other operator contribution.  The
reason is discussed in Appendix.

Using the general relation
\begin{equation}
\varepsilon'=ie^{i(\delta_2-\delta_0)}\left[-\frac{\mbox{Im}A_0}{\mbox{Re}A_0}+
\frac{\mbox{Im}A_2}{\mbox{Re} A_2}\right]\cdot |\frac{A_2}{A_0}|
\end{equation}
and the experimental value $\varepsilon'=(3.78 \pm 0.38)\times 10^{-6}$, we
come to the relation \begin{equation} -\frac{\mbox{Im} \tilde
c_5}{\mbox{Re} \tilde c_5} \left(1- \Omega_{\eta,\eta'} +24.36
\frac{\mbox{Im} \tilde c_7}{\mbox{Im} \tilde c_5} \right)=(1.63 \pm 0.16)
\times 10^{-4},  \end{equation} where $ \Omega_{\eta,\eta'} $ takes into
account the effects of $K^0 \to\pi^0 \eta (\eta') \to \pi^0\pi^0 $
transitions.

Introducing the notation
\begin{equation}
- \frac{\mbox{Im} \tilde c_5}{\mbox{Re} \tilde c_5} =x \frac{\mbox{Im}
\lambda_t}{s_1}, \qquad \frac{24.36}{1-\Omega_{\eta,\eta'}}\cdot
\frac{\mbox{Im} \tilde c_7}{\mbox{Im} \tilde c_5}=-y \end{equation} and
using \begin{equation} (\mbox{Im}\lambda_t)/s_1 \approx s_2 s_3 \sin
\delta =\frac{(1.2 \pm 0.2)\times 10^{-4}}{0.223} \qquad [7]
\end{equation} we can write Eq.(23) for $\Omega_{\eta,\eta'}=0.25 \pm
0.08$ in the form \begin{equation} x(1-y)=0.40\times (1\pm 0.22)
.\end{equation} In the last two equations $s_i$ and $\delta$ are the
parameters of CKM matrix. The Eq.(26) depends on the variables $x$ and
$y$ representing the contribution of QCD penguin and relative
contribution of EWP, respectively.  To move farther, we are enforced to
apply to existing theoretical estimates of one of these variables.

In terms of notations in [8-10]
\begin{equation}
y=\frac{\Pi_2}{\omega}/\Pi_0(1-\Omega_{\eta,\eta'}).
\end{equation}

According to [8] \quad $y\approx 0.3$ and hence $ x=0.57 \pm 0.12$.
But $\varepsilon'/\varepsilon =2.2\cdot 10^{-3}$ or by 30\% is larger
than the experimental value .

In [10], the central value of $y$ is $y \approx 0.5$ and,
consequently, $x=0.80 \pm 0.18$. This result looks as the reliable one.
A very close result $x=0.71\pm 0.27$ can be derived from the result
$ (\varepsilon'/\varepsilon)_{\mbox{EWP}}=(-12 \pm 3)\times 10^{-4}$ \quad
[11] comparing it with the
experimental value $(\varepsilon'/\varepsilon)_{\mbox{exp}}=(16.6 \pm
1.6)\times 10^{-4}$.  But it should be noted that the previous estimates
of $x$ were rather different. In particular, according to [12] \quad
$x=1.4 \pm 0.28$.  An estimate of $x$ can be extracted also from the
papers [13-15] operating with different set of 4-quark operators $Q_i$,
where the combination $C_6 Q_6 $ corresponds to our $c_5 O_5$.  From the
general representation $$ C_6(\mu)=z_6(\mu)+(s^2_2 +s_2s_3 \frac{c_2}{c_1
c_3} \cos \delta)\cdot y_6(\mu) -is_2s_3 \frac{c_2}{c_1c_3} \sin \delta
\cdot y_6(\mu) $$ and the calculated magnitudes of $y_6$ and $z_6$ we find
for $x\approx y_6/z_6: $ \begin{equation} x\approx 2 \quad \mbox{at}\;
\Lambda^{(4)}_{\mbox{QCD}}=0.35\; \mbox{GeV} ,\;\mu=0.8\; \mbox{GeV}, \;
m_t=176\;\mbox{GeV}\quad [13]; \end{equation} \begin{equation} x=2.8 \quad
\mbox{at}\; \Lambda_{\bar {MS}}=0.3\; \mbox{GeV},\; \mu=1\; \mbox{GeV},\;
m_t =130\; \mbox{GeV} \quad [14]; \end{equation} \begin{equation} x=5.5
\quad \mbox{at}\; \Lambda^{(4)}_{\mbox{QCD}}=0.3\; \mbox{GeV},\; \mu=1\;
\mbox {GeV},\; m_t=170\; \mbox{GeV} \quad [15].  \end{equation}

Such difference of the theoretical estimates of $x$ makes very desirable
an investigation of CP-effects in $K^{\pm} \to
\pi^{\pm}\pi^{\pm}\pi^{\mp}$ decays, where, contrary to $K_L \to 2\pi$
decays, the EWP contributions increase CP effects .  \section*{3. Decay
{\bf $K^{\pm} \to \pi^{\pm} \pi^{\pm} \pi^{\mp}$}}

To leading $p^2$
approximation \begin{equation} M(K^+ \to \pi^+(p_1) \pi^+(p_2)
\pi^-(p_3))=\kappa[1+ia_{KM} +\frac{1}{2} g Y(1+ib_{KM})+...],
\end{equation} where \begin{equation} \kappa =G_{\rm F} \sin \theta_{\rm
C} \cos \theta_{\rm C} m^2_K c_0 (3\sqrt{2})^{-1}, \end{equation}
\begin{equation} a_{KM}=\left[\frac{32}{9} \beta \mbox{Im} \tilde c_5 +
4\beta \mbox{Im} \tilde c_7 \left(\frac{3\Lambda^2}{2m^2_K}+2 \right)
\right]/c_0, \end{equation} \begin{equation} b_{KM}=\left[\frac{32}{9}
\beta \mbox{Im} \tilde c_5 +8\beta \mbox{Im} \tilde c_7
\right]/(c_0+9c_4). \end{equation} The last two quantities represent the
imaginary parts produced by the Kobayashi-Maskawa phase $\delta$.
\begin{equation} \frac{1}{2} g=-\frac{3m^2_{\pi}}{2m^2_K}(1+9c_4/c_0)
,\qquad Y=(s_3-s_0)/m^2_{\pi}, \end{equation} \begin{equation}
c_0=c_1-c_2-c_3-c_4 +\frac{32}{9} \beta \mbox{Re} \tilde c_5 =-10.46.
\end{equation}

As the field $K^+$ is the complex one and its phase is arbitrary, we can
replace $K^+$ by $K^+(1+ia_{KM})(\sqrt{1+a^2_{KM}})^{-1}$. Then
\begin{equation}
M(K^+ \to \pi^+(p_1) \pi^+(p_2) \pi^-(p_3))
=\kappa [1+\frac{1}{2}gY(1+i(b_{KM}-a_{KM}))+...].
\end{equation}
Though this expression contains the imaginary CP-odd part, it does not
lead to observable CP effects.  Such effects arise due to interference
between CP-odd imaginary part and the CP-even imaginary part produced by
rescattering of the final pions.  Then
\begin{equation}
M(K^+ \to \pi^+ \pi^+ \pi^-) = \kappa [1+ia +\frac{1}{2}
gY(1+ib+i(b_{KM}-a_{KM}) +...]
\end{equation}
where $a$ and $b$ are corresponding CP-even imaginary parts of the
amplitude. These parts can be estimated to leading approximation in
momenta calculating the imaginary part of the two-pion loop
diagrams with $$ \begin{array}{lll} M(\pi^+(r_2) \pi^-(r_3) \to \pi^+(p_2)
\pi^-(p_3)) =F_{\pi}^{-2}[(p_2+p_3)^2 + (r_2-p_2)^2 -2m^2_{\pi}], \\
M(\pi^0(r_2) \pi^0(r_3) \to \pi^+(p_2) \pi^- (p_3))=F^{-2}_{\pi}
[(p_2+p_3)^2 -m^2_{\pi}], \\ M(\pi^+(r_1) \pi^+(r_2) \to \pi^+(p_1)
\pi^+(p_2))=F^{-2}_{\pi}[(r_1-p_1)^2 +(r_1-p_2)^2 -2m^2_{\pi}]. \end{array}
$$
Then we find:
\begin{equation}
a=  0.12065, \qquad b=0.714.
\end{equation}
Using the definition
$$
|M(K^{\pm} \to \pi^{\pm}(p_1) \pi^{\pm}(p_2) \pi^{\mp}(p_3))|^2 \sim
[1 + g^{\pm}Y +...]
$$
and the results of our calculation
\begin{equation}
|M(K^{\pm} \to  \pi^{\pm}(p_1) \pi^{\pm}(p_2) \pi^{\mp}(p_3)|^2 \sim
[1+\frac{g}{1+a^2}Y\left(1+ab \pm a(b_{KM}-a_{KM})\right)+...]
\end{equation}
we find
\begin{equation}
R_g \equiv \frac{g^+-g^-}{g^+ +g^-}= \frac{a(b_{KM}-a_{KM})}{1+ab}.
\end{equation}
At the fixed above numerical values of the parameters and
$\Omega_{\eta,\eta'}=0.25$ we obtain to leading $p^2$
approximation \begin{equation} \left(R_g \right) _{p^2} =0.030
\frac{\mbox{Im} \tilde c_5}{\mbox{Re} \tilde c_5}(1-14.9 \frac{\mbox{Im}
\tilde c_7}{\mbox{Im} \tilde c_5}) = \\ =-(2.44 \pm 0.44 )\times
10^{-5}x\left(1-\frac{0.13 \pm 0.03}{x}\right).  \end{equation}
\section*{4. The role of {\bf $p^4$} and other corrections}

The
corrections to the result obtained in the conventional chiral theory up to
leading $p^2$ approximation are of two kinds. The first kind corrections
are connected with a necessity to get explanation of the observed
enlargement of $S$-wave $I=0$ $\pi \pi$ amplitude.   The corrections of
the second kind are the $p^4$ corrections. As it was argued in [16], [17],
the corrections of both kinds be properly estimated in the
framework of special linear $U(3)_L \otimes U(3)_R$ $\sigma$ model with
broken chiral symmetry. The above mentioned enlargement of $S$ wave in
this model is originated by mixing between the $\bar q q$ state and the
gluonic state $(G^a_{\mu\nu})^2$ states.  In such a model $$ U=\hat \sigma
+ i \hat \pi $$ where $\hat \sigma$ is $3\times 3$ matrix of scalar
partners of the mesons of pseudoscalar nonet.  The relations between
diquark combinations and spinless fields are as given by eqs.(10) ,(11),
but without  the terms proportional to $\Lambda^{-2}$.  Such contributions
in $\sigma$ model appear from an expansion of the propagators
of the intermediate scalar mesons.   The parameter $\Lambda^2$ in this
model is equal to difference $m^2_{a_0(980)}-m^2_{\pi}$. The
strength of mixing between the isosinglet $\sigma$  meson and
corresponding gluonic state is characterized by the parameter $\xi$.

If the $p^2$ approximation gives \begin{equation}
(\kappa)_{p^2}=1.495 \times 10^{-6}, \qquad (g)_{p^2} =-0.172
\end{equation} instead of \begin{equation}(\kappa)_{\mbox{exp}}=1.92
\times 10^{-6}, \qquad (g)_{\mbox{exp}}=-0.2154 \pm 0.0035, \end
{equation} the corrected values of these CP-even parameters of $K^+ \to
\pi^+ \pi^+ \pi^-$ amplitude are closer or practically equal to the
experimental ones [16]:  \begin{equation} (\kappa)_{(p^2+p^4;\;
\xi=-0.225)}=1.73 \times 10^{-6}, \qquad (g)_{(p^2+p^4;\; \xi
=-0.225)}=-0.21.  \end{equation} More information on the parameter $\xi$
can be found in [16],[17]. The expressions for the corrected $\pi \pi \to
\pi \pi$ amplitudes are presented in [17].

Calculating the CP-even imaginary part of the $ K^{\pm} \to \pi^{\pm}
\pi^{\pm} \pi^{\mp}$ amplitude originated by two-pion intermediate states,
we obtain
\begin{equation}
a(p^2+p^4; \;\xi=-0.225)=0.16265,
\end{equation}
\begin{equation}
b(p^2+p^4;\; \xi=-0.225) =0.762 .
\end{equation}

An estimate of the parameter $a$ can be obtained also without any
calculations using the experimental data on the phase shifts of $\pi\pi$
scattering $\delta^0_0, \,\delta^2_0,\,\delta^1_1$.
According to definition (38) $a$ is a phase at $s_3=s_0$.  The mean value
of the squared energy of $\pi^+ \pi^-$ system is $$ \frac{1}{2}[(p_1+p_3)^2
+(p_2+p_3)^2]=s_0 +\frac{s_0-s_3}{2}.  $$ Consequently, $a$ is a phase
shift of $\pi^+ \pi^-$ scattering at $\sqrt{s}=\sqrt{s_0}$. But the only
significant phase shift at $\sqrt{s}=\sqrt{s_0}$ is $\delta^0_0$.  The
rest phase shifts are very small:  $|\delta^2_0(s_0)|<1.8^{\circ}$ and
$\delta^1_1(s_0) <0.3^{\circ} $ [18].  Then, according to Eq.(38), $a
\approx \tan \delta^0_0(s_0)$, or $a=0.13 \pm 0.05$, if
$\delta^0_0(s_0)=(7.50 \pm 2.85)^{\circ}$ \,[19] and $a=0.148 \pm 0.018$,
if $\delta^0_0(s_0)=(8.4 \pm 1.0)^{\circ}$\, [20].  These results coincide
inside the error bars with the result (46).  The corrected magnitude of
$R_g$ is \begin{equation} \begin{array}{rcl} \left(R_g \right)_{(p^2+p^4;
\; \xi=-0.225)}=0.039\frac{\mbox{Im} \tilde c_5}{\mbox{Re} \tilde c_5}
\left(1-11.95 \frac{\mbox{Im} \tilde c_7}{\mbox{Im} \tilde c_5} \right)=
\\ =- (3.0 \pm 0.5)\times 10^{-5}x\left(1-\frac{0.11 \pm 0.025}{x} \right)
.  \end{array} \end{equation} This result is by 23\% larger in absolute
magnitude than that calculated in the leading approximation.  Therefore,
we come to conclusion that the corrections to the result obtained in the
framework of conventional chiral theory to the leading approximation are
not negligible (23\%), but not so large, as it was declared in [21].
\section*{5. Conclusion}
From Eqs.(22), (26) and (48), it follows that EWP contributions diminish
$\varepsilon'/\varepsilon$ and increase $R_g$. The EWP corrections cancell
one half of the QCD penguin contribution into $\varepsilon'/\varepsilon$
at $x=0.8$ and cancell 80\% of QCD penguin contribution at $x=2$. In both
cases $\varepsilon'/\varepsilon$ is the same.

In the case of $K^{\pm} \to 3\pi$ decays, the {\it direct} influence of
EWP corrections themselves on CP effects is not so crucial as in $K_L \to
2\pi$ decays. But if a cancellation between the contribution of QCD and
electroweak penguins in $\varepsilon'/\varepsilon$ is large, the factor
$x$ in Eq.(48) is also larger than 1. So, for $x=2$ , the predicted $R_g$
must be 2.5 times larger than at $x=0.8$.

Therefore, measuring $R_g$, one obtains a possibility to determine the
true relation between QCD and EWP contributions into CP violation in kaon
decays.

\vspace{15mm}
\hspace{9cm} {\Large {\bf Appendix}}\\

Here we explain why an expansion of the amplitudes originated by
electroweak penguin diagrams begins from the term, independent
of momenta and masses of the pseudoscalar mesons.

The operators $O_{7,8}$ can be expressed in terms of colorless diquark
combinations in the form $$ O_7=-\bar s(1-\gamma_5)u \cdot
\bar u(1+\gamma_5)d -\frac{3}{8}O_5, \qquad O_8=3O_7 .
\eqno ({\rm A}.1) $$
Using eq.(10), we find
$$
O_7=-\frac{F^2_{\pi} r^2}{2}U_{21}U^*_{13} +(\mbox{terms
proportional to} \quad p^2_i ( m^2_i)).
\eqno ({\rm A}.2) $$
Omitting the terms proportional to derivatives of $U$ and taking in
Eq.(12) $a_3=0$, we find for the parity-even transitions
$$
\begin{array}{rcl}
(O_7)^{P-{\rm even}}=-\frac{F^2_{\pi}r^2}{2}\{\pi^-K^+
+\frac{1}{2F^2_{\pi}}
\left[\pi^-(\frac{2\pi_0}{\sqrt{3}}+\frac{2\pi_8}{\sqrt{6}})+K^0
K^-\right] \\ \times
\left[K^+(\frac{2\pi^0}{\sqrt{3}}-\frac{\pi_8}{\sqrt{6}}
+\frac{\pi_3}{\sqrt{2}}) +\pi^+ K^0\right] +...\}.
\end{array}  \eqno ({\rm A}.3)
$$
This expression does not contain the direct contribution to $K^+ \to
3\pi$ decays, but thanks to the term $\pi^- K^+$, the independent of
$p^2_i (m^2_i)$ part of $ K^+ \to 3\pi$ amplitude arises. In $p^2$
approximation
$$
\begin{array}{lll}
<\pi^+(p_1) \pi^+(p_2)
\pi^-(p_3)|O_7|K^+(k)>= \\ =-\frac{F^2_{\pi}r^2}{2(m^2_K-m^2_{\pi})}
\left[\frac{s_1+s_2-2m^2_{\pi}}{F^2_{\pi}}-\frac{s_1
+s_2-m^2_{\pi}-m^2_K}{F^2_{\pi}} \right]=-\frac{r^2}{2},
\end{array}  \eqno ({\rm A}.4)
$$
where $s_i=(k-p_i)^2$ and the first term in the brackets describes the
$\pi^+(k)\to \pi^+(p_1) \pi^+(p_2) \pi^-(p_3)$ transition. The second term
describes the transitions $K^+(k) \to
K^+(p_{1,2})\pi^+(p_{2,1})\pi^-(p_3)$.

Therefore, the operator $O_7$ violates the rule, according to which  an
expansion of the mesonic amplitudes begins from the terms proportional to
$p^2_i(m^2_i)$.

It may seem to one that removing the non-diagonal term $-\frac{r^2}{2}
\pi^- K^+$ from the effective Lagrangian by redefinition of $K^+$ and
$\pi^-$ fields [22], the problem with the constant contribution could
be solved. But this is not so.

In our case, the mass part of the effective Lagrangian contains, in
particular, the combination
$$
-m^2_{\pi} \pi^+\pi^- - m^2_K K^+ K^- -\frac{F^2_{\pi}r^2}{2}(\gamma
K^+\pi^- +\gamma^* K^-\pi^+),  \eqno ({\rm A}.5)
$$
where $\gamma=\sqrt{2}G_{\mbox{F}}\sin \theta_{\mbox{C}} \cos
\theta_{\mbox{C}} c_7$.  The transformations $$ \begin{array}{lll} \pi^-
\to \pi^- +\beta K^-, \qquad K^+ \to K^+ -\beta \pi^+, \\ \pi^+ \to \pi^+
+\beta^* K^+, \qquad K^- \to K^- -\beta^* \pi^- \end{array} \eqno
({\rm A}.6) $$ with $$ \beta = \gamma^* F^2_{\pi}r^2/2(m^2_K-m^2_{\pi})
\eqno ({\rm A}.7) $$ remove the non-diagonal terms in the linear in
$\gamma$ approximation.  But the effective Lagrangian of strong
interaction generates the sum of the amplitudes $$ \begin{array}{rcl}
<\pi^+(p_1)\pi^+(p_2)\pi^-(p_3)|\pi^+(k)>
+<K^+(p_1)\pi^+(p_2)\pi^-(p_3)|K^+(k)> +\\
<K^+(p_2)\pi^+(p_1)\pi^-(p_3)|K^+(k)> \end{array} \eqno ({\rm A}.8)   $$
which after the transformation (A.6) generates the amplitude $$
\begin{array}{lll} <\pi^+(p_1)\pi^+(p_2)\pi^-(p_3)|O_7|K^+(k)>= \\
=-\frac{\beta}{\gamma^*} \cdot \left[\frac{s_1+s_2-2m^2_{\pi}}{F^2_{\pi}}-
\frac{s_1+s_2-m^2_{\pi}-m^2_K}{F^2_{\pi}} \right] =-\frac{r^2}{2}.
\end{array}
\eqno ({\rm A}.9)  $$
We have reproduced the result (A.4). The contribution of the operators
$O_{7,8}$ to the leading approximation does not depend on $p^2_i
(m^2_i)$.\\

\vspace{1cm}
This work is supported in part by Federal Program
of the Russia Ministry of Industry, Science and Technology
No.40.052.1.1.1112.

\vspace{1cm}
\noindent {\large \bf References}

%\begin{thebibliography}{99}
\noindent 1. S.Bertolini
{\it et al.}, Nucl.Phys.B {\bf 514}, 93 (1998). \\  2. T.Hambye {\it et
al.}, Nucl.Phys.B {\bf 564}, 391 (2000). \\  3. M.A.Shifman,
A.I.Vainshtein and V.I.Zakharov, Zh.Eksp.i Teor.Fiz. {\bf
72}, 1277 (1977).  \\  4. W.A.Bardeen, A.J.Buras and J.-M.Gerard,
Nucl.Phys.B {\bf 293}, 787 (1987). \\  5. J.Cronin, Phys.Rev.
{\bf 161}, 1483 (1967). \\ 6. L.B.Okun, {\it Leptons and Quarks}
(North-Holland Publ.Co.  1982) pp.315,323.  \\ 7. A.Ali and D.London
, Eur.Phys.J. C {\bf 18}, 665 (2001). \\ 8. S.Bertolini, J.O.Eeg and
 M.Fabbrichesi, Phys.Rev. D {\bf 63}, 056009 (2001).   \\9.
A.J.Buras and J.-M.Gerard, Phys. Lett. B {\bf 517}, 129 (2001). \\
10. T.Hambye, S.Peris and E.de Rafael,  hep-ph/0305104 v. 2.
\\  11. J.F.Donoghue and E.Golovich,  Phys.Lett. B {\bf
478}, 172 (2000). \\ 12. M.B.Voloshin, preprint ITEP-22. (Moscow 1981).
 \\
13. S.Bertolini {\it et al.}, preprint SISSA 102/95/EP.\\
14. A.J.Buras, M.Jamin and M.Lautenbacher, Nucl.Phys.B {\bf 408}, 209
(1993). \\ 15. S.Bertolini, J.O.Eeg and M.Fabbrichesi, Nucl.Phys. B {\bf
449}, 197 (1995). \\ 16. E.P.Shabalin, Nucl.Phys.B {\bf 409}, 87 (1993).
\\ 17. E.P.Shabalin, Phys.  At. Nucl. {\bf 61}, 1372 (1998).\\ 18.
E.P.Shabalin, Phys.At.Nucl. {\bf 63}, 594 (2000).\\ 19.  L.Rosselet {\it
et al.}, Phys.Rev.D {\bf 15}, 574 (1977). \\ 20.  S.Pislak {\it et al.},
Phys.Rev.Lett. {\bf 87}, 221801 (2001). \\ 21.  A.A.Bel'kov {\it et al.},
Phys.Lett.B {\bf 300}, 283, (1993). \\ 22.  G.Feinberg, P.K.Kabir,
S.Weinberg,  Phys.Rev.Lett.{\bf 3}, 527 (1959).  %\end{thebibliography}

\end{document}